%% file: tocl.tex
\documentclass[12 pt]{llncs}
\usepackage{times}
\usepackage{epsfig}
\usepackage{xspace}
\usepackage{amsfonts}
\input{hmacs}
\input{amacs}

\normalsize
\newcommand\I{\overline{I}}
\newcommand\bv{\overline{b}}
\newcommand\con{`$\land$'}
\newcommand\dis{`$\lor$'}
\newcommand\condis{\land \!\!\!\!\lor}
\begin{document}
\title{
Deciding Disjunctive Linear Arithmetic \\with SAT
\author {Ofer Strichman}
\institute{Computer Science Department, Carnegie Mellon University, 5000 Forbes Avenue, Pittsburgh,
PA 15213-3891, USA\\Email: \email{ofers@cs.cmu.edu}\\
\emph{Keywords}: Program Verification, Decision procedures,  SAT
}
\thanks{This research was supported in part by the Office of Naval Research (ONR) and the Naval
Research Laboratory (NRL) under contract no. N00014-01-1-0796. }
\thanks{An early version of this article
appeared in \cite{S02}.} 
\comment{The main additions in this extended version are:
A theoretical discussion on the advantages of the proposed method
comparing to existing ones, including two new claims in section
\ref{sec:revised}; A thorough experimental comparison of the methods,
including a new benchmark of several hundred randomly generated formulas. }
}

\maketitle

\begin{abstract}
Disjunctive Linear Arithmetic (DLA) is a major decidable theory that is supported by almost all
existing theorem provers. The theory consists of Boolean combinations of predicates of the form
$\Sigma_{j=1}^{n}a_j\cdot x_j \le b$, where the coefficients $a_j$, the bound $b$ and the variables
$x_1 \ldots x_n$ are of type Real ($\mathbb{R}$).  We show a reduction to propositional logic from
disjunctive linear arithmetic based on Fourier-Motzkin elimination.  While the complexity of this
procedure is not better than competing techniques, it has practical advantages in solving
verification problems. It also promotes the option of deciding a combination of theories by reducing
them to this logic. Results from experiments show that this method has a strong advantage over
existing techniques when there are many disjunctions in the formula.
\end{abstract}

\section{Introduction}

Disjunctive Linear Arithmetic (DLA) is a major decidable theory that is supported by almost all
existing theorem provers, and is used frequently when proving infinite state systems. The theory
consists of Boolean combinations of predicates of the form $\Sigma_{j=1}^{n}a_j\cdot x_j \le b$,
where the coefficients $a_j$, the bound $b$ and the variables $x_1 \ldots x_n$ are of type Real
($\mathbb{R}$).

Decision procedures for this theory typically handle disjunctions by `case-splitting', i.e.,
transforming the formula to Disjunctive Normal Form (DNF) and then solving each clause
separately. Naive case-splitting procedures explicitly transform the formula to DNF, and are
therefore very restricted in the size of the formula that they can handle (the number of clauses in
the resulting formula can be exponential in the size of the original formula). More sophisticated
implementations split the formula only `as needed', which increases in many cases the capacity of
these procedures, although there can still be an exponential number of cases to solve. 

Recently a different approach was introduced almost simultaneously by three different groups
\cite{MRS02,ABCKS02,SBD02}. The procedure is based on a combination of a SAT procedure and an arithmetic
solver, and is now implemented by tools such as \tool{cvc}, \tool{mathsat} and
\tool{ics-sat}\footnote{\tool{ics-sat} is the name we call the version of \tool{ics} that works
according to this combined approach. The distinction between the two versions is important in this
article, as \tool{ics} works with case-splitting.}. The procedure works roughly as
follows. The linear predicates are encoded with Boolean variables, and then the encoded Boolean
formula is solved with a SAT solver. If the SAT instance is unsatisfiable, then the procedure
terminates and declares the formulas unsatisfiable. Otherwise, it checks whether the given
assignment is consistent with respect to the linear constraints. This step amounts to solving a
conjunction of predicates or negation of predicates, which is possible by using any number of procedures
(see below). If a satisfying assignment is found, then the procedure terminates and declares the
formula to be satisfiable. Otherwise, it backtracks in order to find a different assignment, while
typically (depending on the specific system) applying a learning mechanism, i.e. adding a Boolean
conflict clause that prevents a repetition of the bad assignment. Although this approach can still
be seen as case splitting, as it still may call the arithmetic solver an exponential number of
times, the learning and pruning power of the SAT solver makes it far more robust than naive
case-splitting methods. We will further discuss the advantages and disadvantages of these techniques
in section \ref{sec:revised}.

The lower-bound complexity of solving each DNF clause, i.e., a conjunction of linear constraints, is
polynomial \cite{K79}. When considering small to medium size problems, as the ones that are
typically encountered in formal verification, the existing polynomial procedures are rarely better
in practice comparing to some exponential methods like Simplex \cite{dantzig63} and the various
variable-elimination techniques. For this reason, as far as we know, no automated theorem prover
uses a polynomial procedure for linear arithmetic. 

The most commonly used method by theorem provers is the Fourier-Motzkin (FM) variable elimination
method \cite{BW94}, which is used in popular tools such as \tool{pvs}\cite{PVSg},
\tool{ics}\cite{FORS01}, \tool{svc}\cite{BDL96}, \tool{imps}\cite{FGT92} and others. We describe the
FM method in detail in section \ref{fourier}.  Although FM has a worst-case super-exponential
complexity, it is popular because it is frequently faster than competing methods for the size of
instances encountered in practice.  Hence, the current practice in solving DLA is to solve, in the
worst case, an exponential number of FM instances. Theoretically this is not the best possible, as
explained above, but experience has showed that for the type of formulas encountered in
verification, it is adequate.

The procedure described in this paper solves one FM instance in order to generate a SAT instance,
and then solves this instance with a standard SAT solver. It has a similar complexity to what we
just described as the common practice, but we expect it to be better in practice because of reasons
that we will later discuss. SAT solvers are generally far more efficient than case splitting in
handling propositional combinations of formulas, although both have the same theoretical
complexity. Propositional SAT checkers apply techniques like {\it learning}, {\it pruning} and {\it
guidance} (`guidance' refers to heuristics for prioritizing the internal steps of the decision
procedure) that can not be easily imitated by case-splitting. We refer the reader to \cite{SSB02}
where an elaborated discussion of this distinction is given. Based on this observation, our
suggested procedure is expected to be more efficient than case-splitting methods in deciding
formulas where the case-splitting itself is the bottleneck of the procedure, i.e., formulas that
their equivalent DNF has many clauses, but each one of them is relatively small.

An efficient reduction of DLA to propositional logic not only enables to (potentially) solve them
faster, but also to integrate them with other theories on the propositional logic level. Many other
decidable theories that are frequently encountered in verification (e.g. bit-vector arithmetic
\cite{J01}) already have such reductions to propositional logic. Solving mixed theories by reducing
them to a common logic facilitates the application of various learning techniques between
sub-expressions that originate from different theories. Furthermore, current popular techniques for
integrating theories such as Nelson-Oppen \cite{NO79} invoke different procedures for deciding each
theory, and propagate equalities between them in order to decide the combined theory. The overhead
of this mutual updating can become significant. This overhead is avoided if only one procedure (SAT
in this case) is used.

The rest of the article is structured as follows. In the next section we briefly describe the FM
method. In section \ref{propositional} we present a propositional version of the same procedure and
explain how it can be used to reduce DLA to SAT. In section \ref{conjunctions} we present a method called `conjunctions matrices',
which is useful for reducing the complexity of the procedure described in section
\ref{propositional}. In section \ref{experiments} we summarize our experiments with this method on
both real examples and random instances.

\section{Fourier-Motzkin Elimination}
\label{fourier}
A linear inequality predicate over $n$ variables has the form $\Sigma_{j=1}^{n}a_j\cdot x_j \le
b$. A conjunction of $m$ such constraints is conveniently described by $C: A\I \le \bv$ where $A$ is
an $m \times n$ real-valued coefficient matrix, $\I = x_1...x_n$ is a vector of $n$ variables, and
$\bv$ is a vector of real-valued bounds. Given a variable order $x_1...x_n$ the FM method eliminates
(existentially quantifies) them in decreasing order. Each variable is eliminated by projecting its
constraints on the rest of the system. The procedure works as follows: at each elimination step, the
list of constraints is partitioned to three segments, according to the sign of the coefficient of
$x_n$ in each constraint. Let $a_{i,n}$ denote the coefficient of $x_n$ in constraint $i$, for $i
\in[1..m]$. 

\begin{minipage}{12 cm}
\vspace{0.1 cm}
The three segments are:
\begin{enumerate}
\item For all $i$ s.t. $a_{i,n}>0$: \hspace{1 cm} \hspace{1.3 cm} $a_{i,n}\cdot x_n\ \le \ b_i - \Sigma_{j=1}^{n-1}a_{i,j}\cdot x_j$
\item For all $i$ s.t. $a_{i,n}<0$: \hspace{1 cm} $\Sigma_{j=1}^{n-1}a_{i,j}\cdot x_j-b_i\ \le \ -a_{i,n}\cdot x_n$ 
\item For all $i$ s.t. $a_{i,n}=0$: \hspace{1 cm} \hspace{0.7 cm}$\Sigma_{j=1}^{n-1}a_{i,j}\cdot x_j\ \le\ b_i$
\end{enumerate}
\vspace{0.1 cm}
\end{minipage}

The first and second segments correspond to upper and lower bounds on $x_n$, respectively. To
eliminate $x_n$, FM replaces each pair of lower and upper bound constraints $L \le c_l \cdot x_n$
and $c_u \cdot x_n \le U$, where $c_l,c_u >0$, with the new constraint $c_u\cdot L \le c_l \cdot U$.
If, in the process of elimination, the procedure derives the constraint $c \le 0$ where $c$ is a
constant greater than 0, it terminates and indicates that the system is unsatisfiable.

Note that it is possible that variables are not bounded from both ends.  In this case it is possible
to simplify the system by removing these variables from the system together with all the constraints
to which they belong. This can make other variables unbounded. Thus, this simplification stage
iterates until no such variables are left.

The FM method can result in the worst case in $m^{2^n}$ constraints, which is the reason that it is
only suitable for a relatively small set of constraints with small number of variables. There are
various heuristics for choosing the elimination order. A standard greedy criteria gives priority to
variables that their elimination produces less new constraints.
\begin{example}
Consider the following formula:
\[
\varphi = x_1 - x_2 \le 0\quad \land \quad x_1 - x_3 \leq 0 \quad \land \quad -x_1 + 2x_3 + x_2\le 0 \quad \land \quad -x_3 \le -1 \] 

\nin The following table demonstrates the elimination steps following the variable order $x_1$,$x_2$,$x_3$:

\begin{center}
\begin{tabular}{|c|c|c|c|}\hline
\ Eliminated\ & Lower & Upper & New        \\
var	   & bound & bound & constraint \\ \hline
$x_1$	   &\ $x_1-x_2\le 0\ $&$\ -x_1+2x_3+x_2\le 0\ $&$2x_3 \le 0$ \\
           &$x_1-x_3\le 0$&$-x_1+2x_3+x_2\le 0$&$\ x_2 + x_3 \le 0\ $ \\ \hline
$x_2$      & \multicolumn{3}{|c|}{ no lower bound}\\ \hline
$x_3$	   & $2x_3\le 0$   &$ -x_3 \le  -1$& $2 \le  0 $\\ \hline
\end{tabular}
\end{center}
The last line results in a contradiction, which implies that this system is unsatisfiable.
\qed
\end{example}

\nin The extension of FM to handle a combination of strict ($<$) and weak ($\le$) inequalities is
simple. If either the lower or upper bound are a strict inequality,
then so is the resulting constraint.

In the next section we present a Boolean version of the FM method.

\section {A Boolean version of Fourier-Motzkin}
\label{propositional}
Given a DLA formula $\varphi$, we now show how to derive a
propositional formula $\varphi'$ s.t. $\varphi$ is satisfiable iff $\varphi'$ is satisfiable. The
procedure for generating $\varphi'$ emulates the FM method.
\begin{enumerate}
\item \label{normalize} Normalize $\varphi$:
\begin{enumerate}
\item \label{eq} Rewrite equalities as conjunction of inequalities.
\item \label{nnf} Transform $\varphi$ to Negation Normal Form (negations are allowed only over
    atomic constraints).
\item Eliminate negations by reversing inequality signs.
\end{enumerate}
\item \label{encode} Encode each inequality $i$ with a Boolean variable $e_i$. Let $\varphi'$ denote
the encoded formula.
\item \label{perform} 
\begin{enumerate}
\item Perform FM elimination on the set of all constraints in $\varphi$, while
assigning new Boolean variables to the newly generated constraints. 
\item At each elimination step, for every pair of constraints $e_i,e_j$ that result in the new
constraint $e_k$, add the constraint $e_i \land e_j \rightarrow e_k$ to $\varphi'$.
\item If $e_k$ represents a contradiction (e.g., $1 \le 0$), replace $e_k$ by \false.
\end{enumerate}
\end{enumerate}
We refer to this procedure from here on as Boolean Fourier Motzkin (BFM).
\begin{example}
\label{ex_fm}
Consider the following formula:
\[\varphi = 2x_1 -x_2 \le 0 \quad \land \quad (2x_2 - 4x_3 \le 0 \quad \lor \quad  x_3 - x_1 \le - 1)\]
By Assigning an increasing index to the predicates from left to right we initially get $\varphi' =
e_1 \land (e_2 \lor e_3)$.

Let $x_1,x_2,x_3$ be the elimination order. The following table illustrates the process of updating
$\varphi'$:

\begin{center}
\begin{tabular}{|c|c|c|c|c|c|}\hline
Elimina- & Lower & Upper & New        & Enco- & Add to $\varphi'$ \\
ted var  & bound & bound & constraint & ding    &		\\ \hline
$x_1$	& $x_3-x_1 \le -1 $ & $2x_1-x_2 \le 0$ & $2x_3-x_2 \le -2$ & $e_4$ & $e_3 \land e_1 \rightarrow e_4$ \\
$x_2$	& $2x_3-x_2 \le -2$ & $2x_2-4x_3 \le 0$& $4 \le 0    $ & \false & $e_4 \land e_2 \rightarrow \mbox{\false} $ \\\hline

\end{tabular}
\end{center}

\nin Thus, the resulting satisfiable formula is:
\[\varphi' = (e_1 \land (e_2 \lor e_3)) \land (e_1 \land e_3 \rightarrow
e_4) \land (e_4 \land e_2 \rightarrow \mbox{\false})\] 
\qed
\end{example}

Example \ref{ex_fm} demonstrates the main drawback of this method. Since in step \ref{encode} we
consider all inequalities, regardless of the Boolean connectives between them, the number of
constraints that the FM procedure adds is potentially larger than those that we would add if we
considered each case separately (where a `case' corresponds to a conjoined list of inequalities). In
the above example, case splitting would result in two cases, none of which results in added
constraints. Since the complexity of FM is the bottleneck of this procedure, this drawback may
significantly worsen the overall run time and risk its usability.

As a remedy, we will suggest in section \ref{conjunctions} a polynomial method that bounds the number of
constraints to the same number that would otherwise be added by solving the various cases
separately.

\subsubsection{Complexity of deciding $\varphi'$.}
The encoded formula $\varphi'$ has a unique structure that makes it easier to solve comparing to a general
propositional formula of similar size. Let $m$ be the set of encoded
predicates of $\varphi$ and $n$ be the number
of variables.
\begin{proposition}
$\varphi'$ can be decided in time bounded by $O(2^{|m|}\cdot |m|^{2^n})$.
\end{proposition}
\begin{proof} SAT is worst case exponential in the number of decided
variables and linear in the number of clauses. The Boolean value assigned to the predicates in $m$
imply the values of all the generated predicates\footnote{Note that the constraints added in step
\ref{perform} are Horn clauses. This means that for a given assignment to the predicates in $m$,
these constraints are solvable in linear time.}. Thus, we can restrict the SAT solver to split only
on $m$. Hence, in the worst case the SAT procedure is exponential in $m$ and linear in the
number of clauses, which in the worst case is $|m|^{2^n}$.
\qed
\end{proof}

\section{Conjunctions matrices}
\label{conjunctions}
Case splitting can be thought of as a two step procedure, where in the first step the formula is
transformed to DNF, and in the second each clause, which now includes a conjunction of constraints,
is solved separately. In this section we show how to predict, in polynomial time, whether a given
pair of predicates would share a clause if the formula was transformed to DNF. It is clear that
there is no need to generate a new constraint from two predicates that do not share a clause.

\subsection{Joining operands}
\label{joining}
We assume that $\varphi$ is normalized, as explained in step \ref{normalize}.  Let $\varphi'_f$
denote the encoded formula after step \ref{encode} and $\varphi'_c$ denote the added constraints of
step \ref{perform} (thus, after step \ref{perform} $\varphi' = \varphi'_f \land \varphi'_c$). All
the internal nodes of the parse tree of $\varphi'_f$ correspond to either disjunctions or
conjunctions. Consider the lowest common parent of two leaves $e_i,e_j$ in the parse tree. We call
the Boolean operand represented by this node the {\em joining operand} of these two leaves and denote
it by $J(e_i,e_j)$.
\begin{example}
\label{ex1} In the formula $\varphi'_f = e_1 \land (e_2 \lor e_3)$, $J(e_1,e_2)=$ \con\ and $J(e_2,e_3)=$ \dis.
\qed
\end{example}

\nin 
For simplicity, we first assume that no predicates appear in $\varphi$ more than once. In section
\ref{handling} we solve the more general case. Denote by $\varphi^D$ the DNF representation of $\varphi$. The
following proposition is the basis for the prediction technique:
\begin{proposition}
\label{share}
Two predicates $e_i,e_j$ share a clause in $\varphi^D$ iff $J(e_i,e_j) =$ \con.
\end{proposition}
\begin{proof}
Recall that $\varphi'_f$ does not contain negations and no predicate appears more than
once. ($\Rightarrow$) Let $node$ denote the node joining $e_i$ and $e_j$, and assume it
represent a disjunction ($J(e_i,e_j)=$\dis). Transform the right and left branches descending from
$node$ to DNF. A disjunction of two DNF formulas is a DNF, and therefore the formula under $node$ is
now a DNF expression. If $node$ is the root or if there are only disjunctions on the path from
$node$ to the root, we are done. Otherwise, the distribution of conjunction only adds elements
to each of the clauses under $node$ but does not join them into a single clause. Thus, $e_i$ and
$e_j$ do not share a clause if their joining operand is a disjunction. ($\Leftarrow$) Again let
$node$ denote the node joining $e_i$ and $e_j$, and assume it represents a conjunction
($J(e_i,e_j)=$\con). Transform the right and left branches descending from $node$ to
DNF. Transforming a conjunction of two DNF sub formulas back to DNF is done by forming a clause for
each sequence of literals from the different clauses. Thus, at least one clause contains $e_i \land
e_j$. Since there are no negations in the formula, the literals in this clause 
remain together in $\varphi^D$ regardless of the Boolean operands above $node$.
\qed
\end{proof}

\nin For a given pair of predicates, it is a linear operation (in the height of the parse tree $h$)
to check whether their joining operand is a conjunction or disjunction. If there are $m$ predicates
in $\varphi$, constructing the initial $m \times m$ {\em conjunctions matrix $M_\varphi$} of
$\varphi$ has the complexity of $O(m^2h)$. $M_\varphi$ is a binary, symmetric matrix, where
$M_\varphi[e_i,e_j] = 1$ if and only if $J(e_i,e_j)=$\con. For example, $M_\varphi$ corresponding
to $\varphi'_f$ of example \ref{ex1} is given by
\[
M_\varphi = \left(
\begin{array}{c|ccc}
      & e_1 & e_2& e_3 \\ \hline 
e_1   & 0   & 1  & 1   \\ 
e_2   & 1   & 0  & 0   \\ 
e_3   & 1   & 0  & 0   \\ 
\end{array} \right)\] 

\nin Given proposition \ref{share}, this
means that these predicates share at least one clause in $\varphi^D$. New entries are added to
$M_\varphi$ when new constraints are generated, and other entries, corresponding to constraints with
non-zero coefficients over eliminated variables, are removed. The entry for a new predicate $e_k$
that was formed from the predicates $e_i,e_j$ is updated as follows: 
\[\forall l \in [1..k-1].\ M_\varphi[e_k,e_l] = M_\varphi[e_i,e_l] \land M_\varphi[e_j,e_l]\] 
This reflects the fact that the new predicate is relevant only to predicates that share a clause with both 
$e_i$ and $e_j$.

\subsection{Handling repeating predicates}
\label{handling}
Practically most formulas contain predicates that appear more than once, in different parts of the
formula. We denote by $e^k_i$, $k \ge 1$ the $k$ instance of the predicate $e_i$ in
$\varphi'$. It is possible that the same pair of predicates has different joining operands,
e.g. $J(e^1_i,e^1_j)=$\con\ but $J(e^1_i,e^2_j)=$\dis. There are two possible solutions to this
problem:
\begin{enumerate}
\item Represent each predicate instance as a separate predicate. 
\item Assign $M_\varphi[e_i,e_j] = 1$ if there exists an instance of $e_i$ and of $e_j$
s.t. $J(e_i,e_j) =$ \con.
\end{enumerate}

The first option leads to a higher complexity of constructing the initial conjunctions matrix,
because it is determined by the number of predicate instances rather than the number of unique
predicates. More specifically, if $m'$ denotes the number of predicate instances, then the
complexity of constructing the initial matrix $M_\varphi$ is $O(m'^2h)$.

\nin The second option has a more concise representation, but may result in redundant constraints, as the
example below demonstrates.
\begin{example}
Let $\varphi'_f = e_1 \land (e_2 \lor e_3) \lor (e_2 \land e_3)$.  According to option 2, $\varphi'$
contains only three predicates $e_1 \dots e_3$ and therefore $M_\varphi$ is a $3 \times 3$ matrix
with an entry `1' in all its cells. Thus, $M_\varphi$ does not contain the information that the
three predicates never appear together in the same clause, which potentially results in
redundant constraints.  \qed
\end{example}

\nin Conjunctions matrices can be used to speed up many of the other decision procedures that were
published in the last few years for subset of linear arithmetic
\cite{GSZAS98,BV00,BGV99,BGV01,PRSS99,SSB02}. We refer the reader to a technical report \cite{S02tech}
for a detailed description of how this can be done.

\subsection{A revised decision procedure and its complexity}
\label{sec:revised}
Given the initial conjunctions matrix $M_\varphi$, we now change step \ref{perform} as follows:


\begin{enumerate}
\item[\ref{perform}.] \label{perform2} 
\begin{enumerate}
\item Perform FM elimination on the set of all constraints in
$\varphi$, while assigning new Boolean variables to the newly generated constraints.
\item \label{ateach2} At each elimination step consider the pair of constraints $e_i,e_j$ only if $M_\varphi[e_i,e_j] = 1$. In this case let $e_k$ be the new predicate.
\begin{enumerate}
\item Add the constraint $e_i \land e_j \rightarrow e_k$
to $\varphi'$.
\item If $e_k$ represents a contradiction (e.g., $1 \le 0$), replace $e_k$ by \false.
\item Otherwise update $M_\varphi$ as follows: \\ $\forall l \in [1..k-1].\ M_\varphi[e_k,e_l] =
M_\varphi[e_i,e_l] \land M_\varphi[e_j,e_l]$.

\end{enumerate}
\end{enumerate}
\end{enumerate}

\nin 
\newcommand\bfm{$bfm$\xspace}
\newcommand\split{$split$\xspace}
\newcommand\comb{$comb$\xspace}

The main difference between this procedure and the previous one is that now step \ref{perform}(b) is
restricted to pairs of predicates that are conjoined in the DNF of the formula.

Given the revised procedure, we now compare the number of constraints that it generates comparing to
the case-splitting methods, and the combined SAT/FM method \cite{MRS02,ABCKS02,SBD02} that was
described in the introduction. Let \bfm, \split and \comb be the number of constraints that are
generated by these three techniques, respectively.

\begin{nclaim}
For unsatisfiable formulas, BFM generates less or equal number of constraints to the accumulated number of constraints that are
generated by case splitting (\bfm $\le$ \split).
\end{nclaim}
This claim can be easily justified with the observation that due to conjunctions matrices, no
constraint is generated in BFM that is not a resolvent of two constraint in a DNF clause. This means
that the same resolvent is generated by case-splitting methods. In satisfiable instances, the number
of constraints generated by case splitting depends on the location of the first satisfiable
clause. While case splitting terminates after finding the first such clause, \bfm generates all
constraints.

\begin{nclaim}
In most cases in which the formula is unsatisfiable,  \bfm$ \ll$ \split. 
\end{nclaim}
The reason for the big difference between the two procedures is that constraints that are repeated
in many separate cases resolve in a single new constraint in BFM. For example, naive case splitting
over the formula $\varphi' = e_1 \land e_2 \land (e_3 \lor e_4)$ generates the resolvent of $e_1$
and $e_2$ twice, while BFM only generate it once\footnote{Smarter implementation of case splitting
can identify, in this simple example, that the resolvent has to be generated once. But in the
general case redundant constraints can be generated.}. As states above, the comparison of the two
methods is harder in the case of satisfiable formulas, since the number of constraints generated by
case splitting procedures depends on the location of the first satisfiable clause.

The value of \comb is harder to compare to \bfm and \split, because in practice it strongly depends
on the success of the heuristics in the SAT procedure to prune the search space. By guiding the
search, the SAT solver may eventually call the arithmetic procedure for only a small subset of the
possible combinations of predicates. In the worst case, however, \comb can be larger than \split, because it may
generate resolvents of constraints that belong to different DNF clauses (adding conjunctions
matrices to this method can solve this problem. Such an optimization was not described, though, in
the literature \cite{MRS02,ABCKS02,SBD02}). 

Conjunctions matrices is not the only reason for the potentially larger number of constraints that
are generated by the SAT/FM combined procedure. Unlike BFM, this algorithm may generate the same
constraint more than once. Such repeated resolution can occur, for example, if a pair of consistent
predicates appear in many satisfying assignments. When each of these assignments is checked for
consistency, the resolvent of this pair is potentially regenerated. Although saving this information
in a hash table may save some of this repeated work, it may introduce a new source of complexity
because of the possibly exponential number of resolvents. 

A third source for a large number of redundant constraints in the combined procedure, which does not
occur in \tool{bfm}, is the following. Given a set of predicates $p_1 \ldots p_n$, assume that only
$p_1$ and $p_2$ are contradictory. Once the conflict in the set $p_1\ldots p_n$ is identified, a
conflict clause of size $n$ is added, which prevents a repetition of this assignments. This clause
does not, however, prune the other $2^{n - 2} - 1$ contradictory assignments to this set. There are
several solutions to this problem, all of which are either computationally expensive or not
optimal. \tool{cvc} tries to overcome this problem by identifying a small (yet not necessarily
minimal) subset of these literals that actually cause the conflict. In our example, ideally it
identifies that $p_1$ and $p_2$ alone cause the conflict. Consequently it adds a conflict clause of
size two, pruning away the redundant assignments as well as the corresponding resolvents and
conflict clauses. The \tool{ics-sat} tool \cite{MRS02} copes with this problem by following a
trial-and-error approach, in which in each step it tries to remove a predicate and see whether the
conflict still occurs. If the answer is affirmative - it removes the reference to this predicate from the
conflict clause. The success of this approach naturally depends on the order in which the predicates
are removed, and in general does not detect a minimal subset.

\comment{
\section{A generalized framework}
The proposed method is a very simple example of a more broad framework
of converting decision procedures for disjunctive theories to
SAT. With this framework, every procedure that is based on generating and solving a
sequence of problems $\varphi_0 \ldots \varphi_n$ (where $\varphi_0$
is the original problem and for all $i > 0$ $\varphi_i$ is satisfiable
only if $\varphi_{i-1}$ is satisfiable) can be extended to handle a
Boolean combination of the theory through SAT. The framework is the
following: 
\begin{enumerate}
\item  Let $S, |S| = n$ denote the set of existing distinct atoms in the formula. Encode each atom $a_i \in S$ (e.g. a predicate) with a Boolean variable $e_i$. 
\item Use the decision procedure to decide $\bigwedge_{i=0}^{n-1}$.
\item If a new subformula is derived from
a subset $s$ of $S$, add an implication $\wedge_{s_i \in s}s_i \rightarrow s_{n+1}$ where
$s_{n+1}$ is the encoding of the new subformula. $s_{n+1}$ is replaced by the constant \false if a
contradiction is derived.
\item Solve the resulting SAT instance.
\end{enumerate}

In the FM case, each $\varphi_i$ corresponds to a projection of the problem to a reduced set of
variables, hence $n$ is bounded by the number of variables in the original formula $\varphi_0$. In
the case of equality formulas (Boolean combination of equalities), the procedure of Bryant et
al. \cite{Bryant} can be seen as an instance of this framework as well: 
}

\section{Experiments}
\label{experiments}
To test the efficiency of BFM, we implemented a tool called \tool{bfm} on top of \tool{porta} \cite{porta}. We
then randomly generated formulas in 2-CNF style (that is, a 2-CNF where the literals are linear
inequalities) with different number of clauses and variables. The coefficients were chosen
randomly in the range $-10..10$. The time it takes to generate the SAT instance with \tool{bfm} is
summarized in \Figr{table}. The time it takes Chaff \cite{chaff01} to solve each of the instances
that we are able to generate is relatively negligible. Normally it is less than a second, with the
exception of 3 instances that take 10-20 seconds each to solve. All experiments were run on a 1.5
GHz AMD Athlon machine with 1.5 G memory, on top of Linux.

\begin{figure}
\begin{center}
\begin{tabular}{|c||c|c|c|c|c|c|c|c|}\hline
      & \multicolumn{8}{c|}{\# clauses} 	 \\ \hline
\# vars  &	{\ \bf 10}&{\ \bf 30}&{\ \bf 50}&{\ \bf 70}&{\ \bf 90}&{\bf 110}&{\bf 	130}&{\bf 150} \\ \hline \hline
{\bf 10}&	\e &	0.2&	0.2&	1.1&	56&	103&	208&	254 \\ \hline
{\bf 30}&	\e &	0.1&	0.2&	2.5&	61.1&	68&	618&	\E \\ \hline
{\bf 50}&	0.1&	0.1&	0.2&	0.3&	4.9&	8&	173&	893 \\ \hline
{\bf 70}&	0.1&	0.2&	0.2&	0.4&	13.4&	108&	\E&	\E \\ \hline
{\bf 90}&	0.2&	0.2&	0.3&	0.3&	0.5&	1&	14&	181 \\ \hline
{\bf 110}&	0.3&	0.3&	0.5&	8.2&	396&	594&	\E&	\E \\ \hline
{\bf 130}&	0.3&	0.3&	0.4&	0.7&	2.9&	195&	2658&	\E \\ \hline
{\bf 150}&	0.2&	0.3&	0.5&	0.8&	18.4&	334&	1227&	\E \\ \hline
\end{tabular}
\caption{Time, in seconds, required for generating a SAT instance for
random 2-CNF style linear inequalities with a varying number of 
clauses and variables. `\E' indicates running time exceeding 2 hours. }
\label{fig:table}
\end{center}
\end{figure}
We also ran these instances with \tool{ics} and \tool{cvc}. \tool{ics} solves these type of formulas
with FM combined with case-splitting, while \tool{cvc} implements a combined SAT/FM procedure, as
described in the introduction. Both tools can solve only one of these instances (the 10 x 10
instance) in the specified time bound. They either run out of memory or out of time in all other
cases. This is not very surprising, because in the worst case $2^c$ separate cases need to be
solved, where $c$ is the number of clauses. 

The CNF style formulas are harder not only for  \tool{ics} and
\tool{cvc}, but also for \tool{bfm} because they make conjunctions
matrices ineffective. Each predicate in $\varphi$ appears with all
other predicates in some clause of $\varphi^D$, except those
predicates it shares a clause with in $\varphi$. Thus, almost all the
entries of $M_\varphi$ are equal to `1'. In general, conjunctions
matrices only prevent \bfm from adding redundant constraints, and in
CNF formulas only little redundancy is created in the first place. In
order to check the effectiveness of these matrices and experiment with
a larger set of formulas, we ran another batch of examples, where this
time the Boolean connectives (conjunction or disjunction) between the
linear constraints is chosen randomly. That is, a formula with $n$
variables and $m$ clauses has the form ${\condis}_{1 \ldots m}(p(n)
\ {\condis}\ p(n))$ where $\condis$ denotes either a conjunction or a
disjunction, and $p(n)$ is a linear predicate with $n$ variables and randomly chosen coefficients.  For each cell in the table of figure
\ref{fig:table-new} we generated six random instances (a total of 384
random formulas). The numbers in the table represent the average time
it takes to generate the SAT instance with BFM \emph{without}
conjunctions matrices. For comparison, the time it takes to generate
the corresponding SAT instances \emph{with} conjunctions matrices is
almost negligible (a few seconds to generate the entire set). The
reason for this performance can be attributed to the random
construction which apparently results in very few concurrent
constraints.  As before, solving the generated SAT formulas does not
consume a significant amount of time. We also ran \tool{cvc} on this
batch of examples. \tool{cvc} can solve 18 formula out of the 384
rather rapidly (the longest took about three minutes), but exceeds the
time bound or, more frequently, runs out of memory in all other cases.

There are several interesting things to note about the results in
figure \ref{fig:table-new}. First, the results tend to be worse when
the ratio between the number of clauses to number of variables is
high. This is not surprising because FM is sensitive to the product of
upper and lower bounds on each variable. The higher the ratio is, the
larger this product is on average. Second, although not listed here,
there seems to be a very large variance between the different samples,
in particular when the formulas are large. For example, the standard
deviation of the results in each of the cells in the right-most column
is around 400. The reason for these extreme differences is not the
different Boolean structures (to which BFM is insensitive if
conjunctions matrices is inactive), rather it is the different number
of lower and upper bounds on each variable, which is determined by the
randomly selected sign of the coefficients.

\begin{figure}
\begin{center}
\begin{tabular}{|c|c|c|c|c|c|c|c|c|}	 \hline
   & \multicolumn{8}{c|}{\# clauses} 	 \\ \hline
\# vars  &	{\ \bf 10}&{\ \bf 30}&{\ \bf 50}&{\ \bf 70}&{\ \bf
90}&{\bf 110}&{\bf 	130}&{\bf 150} \\ \hline \hline
{\bf 10}	& \e	& \e	& 0.1	& 0.5	& 2.4	& 2.8	& 385.0	& 719.8	\\ \hline
{\bf 30}	& \e	& \e	& 0.1	& 0.7	& 0.3	& 174.2	& 534.4	& 672.0	\\ \hline
{\bf 50}	& \e	& \e	& 0.2	& 1.6	& 3.9	& 114.3	& 393.3	& 696.0	\\ \hline
{\bf 70}	& \e	& \e	& 0.2	& 4.2	& 1.2	& 10.2	& 542.3	& 446.1	\\ \hline
{\bf 90}	& \e	& 0.1	& 0.1	& 0.4	& 0.6	& 285.2	& 103.4	& 425.4	\\ \hline
{\bf 110}& \e	& 0.1	& 0.2	& 0.7	& 0.3	& 8.27	& 107.4	& 171.0	\\ \hline
{\bf 130}& \e	& 0.1	& 0.2	& 0.7	& 0.7	& 1.37	& 13.8	& 166.6	\\ \hline
{\bf 150}& \e	& 0.1	& 0.2	& 0.3	& 0.5	& 0.55	& 0.7	& 0.8	\\ \hline
\end{tabular}
\caption{Average time, in seconds, required for generating a SAT
instance for a formula with random Boolean structure, without
conjunctions matrices. With conjunctions matrices the time is almost
negligible.} 
\label{fig:table-new}
\end{center}
\end{figure}

\begin{figure}
\begin{center}
\begin{tabular}{|c|c|c|c|c|}\hline
\bf{Source}		& \bf{Instance}& \ \tool{bfm}\	& \ \tool{ics}\  &\ \tool{cvc}\ 	\\ \hline
Hardware 	& 1-- 5	& \e	& \e	& \e	\\ \cline{2-5}
designs	 	& 6-7	& \e	& \E	& \e	\\ \hline
Scheduling 	& 1--2	& \e	& \e	& \e	\\ \cline{2-5}
problems    	& 3	& 90	& \E	& \e	\\ \cline{2-5}
		& 4  	&  3    & 952   & 221 	\\ \hline 
Timed       	& 1-2 	& \e    & \e    & \e  	\\ \cline{2-5}
Automata  	& 3    	& \e    & 35    & \e 	\\ \hline \hline
Random 		& 1 	& \E	& 2	& \E 	\\ \cline{2-5}
(Conjoined) 	& 2  	& \E	& 7	& \E 	\\ \hline
\end{tabular}
\caption{ Results achieved by the three tested solvers on several realistic examples from different origins. `\E' indicates running time exceeding 2 hours. }
\label{fig:table2}
\end{center}
\end{figure}
Next, we ran \tool{bfm}, \tool{ics} and \tool{cvc} on several real examples. The results, which are
not as conclusive as with the random instances (many of them can be solved easily by all three
tools), are summarized in figure \ref{fig:table2}. As in the random
instances, here too there seems to be an extreme variation in the
performance of the tools with respect to the different formulas, which can probably be attributed to
the FM method. If the number of constraints starts to grow exponentially, it is typically impossible
to solve the instance in a short time. The examples shown in the table are the following. The first
batch includes seven formulas resulting from symbolic simulation of hardware designs. The second
batch includes four formulas resulting from scheduling problems.  The third batch of examples
contains three standard timed-automata verification problems, namely the verification of a railroad
crossing controller. The first three sets of examples consist of a Boolean combination of
\emph{separation predicates} rather than full linear arithmetic, i.e. predicates of the form $x < y
+ c$, where $c$ is a constant. This is obviously a special case of linear arithmetic.  We also
examined two standard \tool{ics} benchmarks, `linsys-035' and `linsys-100', which consist of 35 and
100 variables and linear inequalities, respectively. The results corresponding to these examples
appear as the last batch in the table. Note that while \tool{ics} solves these instances in a few
seconds, both \tool{bfm} and
\tool{cvc} cannot solve them in the specified time limit. The reason for this seemingly inconsistency is that the \tool{ics} benchmark
formulas consist of a conjunction of linear equalities, and therefore no case splitting is
required. The better performance of \tool{ics} can be attributed to the higher quality of
implementation of FM comparing to that of \tool{porta}, on top of which \tool{bfm} is built, and \tool{cvc}.

Our conclusion from the experiments is that the advantage of \tool{bfm}, as stated in the
introduction, is in solving formulas that have a large number of disjunctions and hence are hard for
any method that is based on solving the various cases separately. The results in figures
\ref{fig:table} and \ref{fig:table-new} prove this observation. The results shown in figure \ref{fig:table2}, however, are
not conclusive. \tool{bfm} has recently been integrated in the theorem prover C-\tool{prover}
\cite{K02}, which means that in the long run additional data concerning the performance of this
technique when solving real verification problems will be gathered.

Finally, as direction for future research, we note that since both DLA and SAT are NP-complete,
there is no complexity argument to rule out the option of finding a polynomial reduction of DLA to
SAT. Finding such a reduction will enable to solve larger formulas than can be solved by \tool{bfm}.

\bibliographystyle{plain} 
\bibliography{biblio}

\end{document}

%% file: hmacs.tex
\usepackage{xspace}

\newcommand\false{{\sc false}\xspace}
\newcommand\tool[1]{{\sc #1}\xspace}
\newcommand\nin{\noindent}

\newcommand\e {$< 1$}
\newcommand\E {*}

\newcommand\corn{\ \lower 5pt\hbox{\hskip 4pt\vrule width 3pt height 1pt depth
  1pt\hbox{\vrule width 2pt height 4.5pt depth 1pt}}}        

\newtheorem{nclaim}{Claim}

\newcommand\Figr[1]{Fig.~\ref{fig:#1}}

\newcommand\comment[1]{}


%% file: amacs.tex
\newcommand\bEx{\begin{example}\rm}
\newcommand\eEx{\hfill\qed\end{example}}

\renewcommand\xi{$x_i$}
